\documentclass{article}

\usepackage{PRIMEarxiv}
\usepackage[utf8]{inputenc}
\usepackage[T1]{fontenc}
\usepackage{url}
\usepackage{booktabs}
\usepackage{amsmath}
\usepackage{amsfonts}
\usepackage{amssymb}
\usepackage{nicefrac}
\usepackage{microtype}
\usepackage{fancyhdr}
\usepackage{graphicx}
\usepackage{float}
\usepackage{makecell}
\usepackage{array}
\usepackage{tabularx}
\usepackage{subcaption}
\usepackage{algorithm}
\usepackage{algpseudocode}
\usepackage{listings}
\usepackage{xcolor}
\usepackage[nocompress]{cite}
\makeatletter
\renewcommand{\citepunct}{,\penalty\@m}
\makeatother
\usepackage{hyperref}
\hypersetup{
  hidelinks,
  pdftitle={Ethereum NFT Smart Contracts: Knowledge-Guided Vulnerability Detection with LLM and Code Slicing},
  pdfauthor={Deyu Yang, Rundong Wei, Xiaoqi Li},
  pdfkeywords={Ethereum, NFT, vulnerability detection, LLM, code slicing}
}

\graphicspath{{./}}
\definecolor{codebg}{RGB}{248,248,248}
\definecolor{codeframe}{RGB}{205,205,205}
\title{Ethereum NFT Smart Contracts: Knowledge-Guided Vulnerability Detection with LLM and Code Slicing}

\author{
  \begin{tabular}{ccc}
    \textbf{Deyu Yang}\thanks{These authors contributed equally to this work.} & \textbf{Rundong Wei}\textsuperscript{*} & \textbf{Xiaoqi Li} \\
    \textnormal{Hainan University} & \textnormal{Hainan University} & \textnormal{Hainan University} \\
    \textnormal{Haikou, China} & \textnormal{Haikou, China} & \textnormal{Haikou, China} \\
    \texttt{3295773553@qq.com} & \texttt{weirundong2026@hainanu.edu.cn} & \texttt{csxqli@ieee.org}
  \end{tabular}
}

\begin{document}
\maketitle

\begin{abstract}
Ethereum non-fungible tokens (NFTs) implement ownership, transfer, authorization, and metadata operations through smart contracts, making contract vulnerabilities a direct risk to digital assets. Existing static analyzers provide efficient rule-based screening but can struggle with application-specific logic, whereas unconstrained large language model analysis may be distracted by irrelevant code or produce inconsistent outputs. We present a vulnerability-detection method that combines vulnerability-focused code slicing, an ERC-721-oriented knowledge base, and constrained DeepSeek analysis. Regular-expression patterns locate candidate statements for reentrancy, integer overflow or underflow, and timestamp dependence. A structure-aware context-window algorithm then extracts line-numbered code slices. DeepSeek analyzes each slice using explicit decision rules and a fixed output schema, and the resulting records support automated batch processing. On 450 NFT contract samples, the full configuration produced 437 positive labels, corresponding to a reported positive-label rate of 97.1\%. Removing the external knowledge base reduced this rate to 87.11\%, while analyzing complete contracts without the knowledge base reduced it to 73.78\%. These results indicate that focused code context and domain constraints materially affect the detector's reported output.
\end{abstract}


\section{Introduction}

Smart-contract security research increasingly combines program analysis, learned representations, and domain-specific reasoning. Recent studies have examined cross-contract exploit detection in decentralized finance and interaction-aware bytecode analysis \cite{LiDeFi2025,Yao2024,LiInteraction2025}. Other work investigates BERT-based bytecode detection, multimodal atomicity analysis, security concerns in non-Ethereum contract ecosystems \cite{Bu2025,He2023,LiAtomGraph2025,Eshghie2023,Wu2025}, exploitable-pattern taxonomies \cite{Vidal2023,Zhang2023,item41}, and LLM-powered hybrid auditing \cite{Yuan2025,Wei2025}. Broader blockchain-security studies examine system hardening, cryptographic foundations, and Ethereum-based security analysis \cite{item17,Wu2024,Li2022}. Collectively, these studies suggest that contract behavior cannot always be understood from isolated syntax because relevant evidence may span calls, bytecode structure, platform semantics, and execution context.

This problem is particularly important for Ethereum non-fungible token (NFT) contracts. ERC-721 contracts govern unique assets in identity, ownership, exchange, and application-specific workflows. Research on NFT ecosystems has examined their socio-technical development, public-key infrastructure, dispute resolution, and financial assurance \cite{Cao2025,item43,Schmitz2022,Ariza2024}. Other studies address industry applications, on-chain NFT scams, and formal verification \cite{Tharun2023,LiNFTScams2025,item51}. Security defects in these contracts can invalidate transfers, corrupt token accounting, or expose assets. We focus on three representative classes: reentrancy, integer overflow or underflow, and timestamp dependence.

Detecting these vulnerabilities is difficult because neither broad static screening nor unrestricted model inference is sufficient on its own. Pattern-based tools are fast and transparent, but matching a transfer, arithmetic operator, or timestamp reference does not establish exploitability. Smart-contract security research has therefore examined attacks and protections, fuzzing, testing, and vulnerability taxonomies \cite{item17,Wu2024,Vidal2023,Barboni2022,item41}. Complementary detection studies use dual-view representations, general vulnerability analysis, exploitable-bug studies, fine-grained analysis, and knowledge-graph-assisted LLMs \cite{Yao2024,He2023,Zhang2023,ShenIntelliCon2023,Wang2023,LiCKGLLM2025}. Other work studies denial-of-service defenses, which remain outside our three-class threat scope \cite{ZhangDoS2025}. More recent LLM-based systems use prompting, planning, benchmarking, or secure-generation constraints to reason about contract behavior \cite{Yuan2025,Liu2025,Peng2025,Wei2025}. However, sending complete contracts to an LLM introduces irrelevant context, whereas free-form prompts make the results difficult to parse and compare.

To address this problem, we combine high-recall candidate screening with focused semantic analysis. We first locate statements associated with the three target vulnerability classes and then extract line-numbered code slices through a context window that expands toward nearby function boundaries. DeepSeek analyzes each slice using an ERC-721-oriented knowledge base, a closed set of allowed labels, and a structured output format. Static patterns reduce the search space, while the LLM evaluates the surrounding logic instead of treating every match as a vulnerability.

We evaluated the method on 450 NFT contract samples. The full configuration produced 437 positive labels and 13 \texttt{None} labels, giving a reported positive-label rate of 97.1\%. In two ablations, this rate decreased to 87.11\% without the external knowledge base and to 73.78\% when complete contracts were analyzed without the knowledge base. These results support the narrower conclusion that code slicing and domain constraints affect the detector's output. They do not establish general accuracy because the evaluation did not use independently adjudicated ground truth.

We make the following contributions:
\begin{enumerate}
  \renewcommand{\labelenumi}{(\arabic{enumi})}
  \item We formulate a focused detection workflow for reentrancy, integer overflow or underflow, and timestamp dependence in Ethereum NFT smart contracts.
  \item We use vulnerability patterns and a structure-aware context window to convert complete contracts into line-numbered, vulnerability-focused code slices.
  \item We constrain DeepSeek with ERC-721 domain knowledge, explicit decision rules, and a fixed output schema that supports automated batch parsing.
  \item We report batch results, runtime measurements, vulnerability distributions, and ablations for the external knowledge base and code-slicing components.
\end{enumerate}

\section{Related Work}

\paragraph{Smart-contract foundations and applications.}
Smart contracts have been studied as programmable agreements \cite{Kaur2023,Ma2022,Sayal2023}, decentralized-finance components \cite{John2023}, and blockchain application infrastructure \cite{YangBlockchain2025,Taherdoost2023,GaoEthereum2025,Lin2022}. Their adoption also raises economic, legal, organizational, and domain-specific questions in banking, construction, trade finance, and Sharia-compliant applications \cite{Kostyuk2025,Yu2023,Ye2022,Wang2022,Fitri2023}. Other work analyzes the conceptual, contractual, and legal status of smart contracts and the relationship between executable code and conventional agreements \cite{Gamper2023}. These studies establish the broader application context but do not directly address vulnerability-focused NFT code analysis.

\paragraph{Languages, specifications, and trust.}
Research on smart-contract languages and standardization examines how programming models, declarative specifications, and reusable components affect contract correctness \cite{Bartoletti2024,Capocasale2022,Chen2022,Park2023}. Knowledge graphs and secure-by-design dataflow models have been used to improve semantic interoperability and implementation discipline \cite{VanWoensel2024,Casale-Brunet2023}. Other studies emphasize that security depends on ownership, protocol assumptions, explainability requirements, and risk-governance choices rather than code alone \cite{Lamby2023,Alshahrani2023,AlGhanmi2024}. Security protocols may benefit from smart contracts, but only when their assumptions and interfaces are explicit \cite{ZhouCryptoReview2025,Li2022}. Our method operates at a narrower layer by using domain knowledge to constrain code-level vulnerability analysis.

\paragraph{Vulnerability taxonomies, testing, and fuzzing.}
Smart-contract security work has developed vulnerability taxonomies, testing strategies, metamorphic testing, and snapshot-based fuzzing \cite{Vidal2023,Barboni2022,Li2023,Shou2023,DingDefense2025}. Surveys of fuzzers show substantial variation in input generation, feedback mechanisms, and vulnerability coverage \cite{Wu2024}. Cross-chain monitoring and comparative studies further demonstrate that practical tools and academic analyses often differ in their assumptions and supported threat models \cite{Eshghie2023,Salzano2025}. Recent work also argues that contract security should extend beyond detection toward explanation, mitigation, and lifecycle support \cite{Abdelaziz2026}. In contrast, we do not introduce a new fuzzer or repair system; we narrow the LLM input with static patterns and code slices.

\paragraph{Learning-based and LLM-based detection.}
Learning-based methods combine source-code, graph, semantic, and structural views to improve vulnerability detection \cite{Yao2024,Wang2023,Bu2025,LiInteraction2025}. LLM-based auditing extends this direction through prompt-guided analysis, dynamic planning, benchmark evaluation, and security-aware generation \cite{Yuan2025,Wei2025,Liu2025,Peng2025,LiLLMHybrid2025}. Adjacent AI-security studies address facial recognition and autonomous tool-invoking agents \cite{LiFacial2025,LiOpenClaw2026}; these settings motivate broader security analysis but remain outside our contract-level detector. Our method addresses a bounded technical task by combining vulnerability-focused slicing with a fixed NFT knowledge base and a deterministic response schema; it does not train or fine-tune a new model.

\section{Background and Threat Scope}

\subsection{Ethereum NFT Smart Contracts}

Ethereum distinguishes externally owned accounts from contract accounts whose code executes in response to transactions. ERC-721 standardizes ownership queries, transfers, approvals, and metadata for unique token identifiers. This standard interface improves interoperability, but each implementation can still introduce errors in state updates, arithmetic, authorization, callbacks, and environmental checks.

\subsection{Target Vulnerabilities}

Table~\ref{tab:vulnerabilities} summarizes the three vulnerability classes evaluated in this study. The scope is deliberately limited; other important classes such as access-control failures, unchecked calls, denial of service, oracle manipulation, and upgradeability risks are outside the current detector.

\begin{table}[H]
\centering
\caption{Target vulnerability classes and representative high-risk contexts.}
\label{tab:vulnerabilities}
\small
\begin{tabularx}{\textwidth}{>{\raggedright\arraybackslash}p{0.20\textwidth} X X}
\toprule
Class & Core condition & Representative contexts \\
\midrule
Reentrancy & An external call occurs before the relevant state update and no effective reentrancy guard prevents repeated entry. & Token transfers, withdrawals, purchase functions, and \texttt{onERC721Received} callbacks. \\
Integer overflow/underflow & Arithmetic exceeds the valid range under legacy or explicitly unchecked behavior. & Minting, burning, balance accounting, supply updates, and exponentiation. \\
Timestamp dependence & A miner-influenced timestamp is used as the decisive basis for a security-sensitive outcome. & Auctions, fundraising windows, blind-box opening, and whitelist activation. \\
\bottomrule
\end{tabularx}
\end{table}

Static pattern matching can find candidate occurrences of these operations, but semantic context is necessary. A transfer can be safe when state is updated first; arithmetic can be checked by the language or a library; and timestamps can be used for noncritical logging. The LLM stage therefore receives the surrounding function context and is instructed to reject matches that do not satisfy the vulnerability definition.

\section{Method}

\subsection{Overall Architecture}

\begin{figure}[!t]
\centering
\begin{subfigure}[t]{0.92\textwidth}
  \centering
  \includegraphics[width=0.82\linewidth]{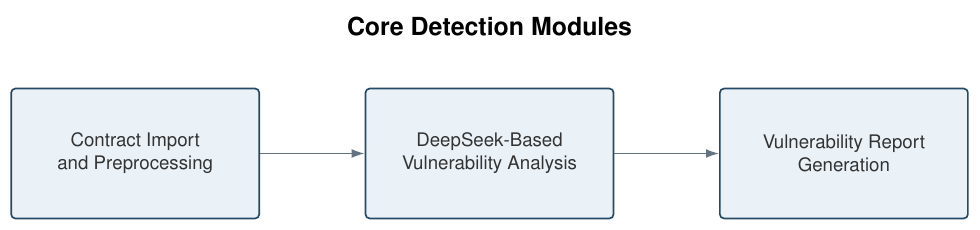}
  \caption{Functional modules.}
\end{subfigure}

\vspace{0.7em}
\begin{subfigure}[t]{0.96\textwidth}
  \centering
  \includegraphics[width=0.96\linewidth]{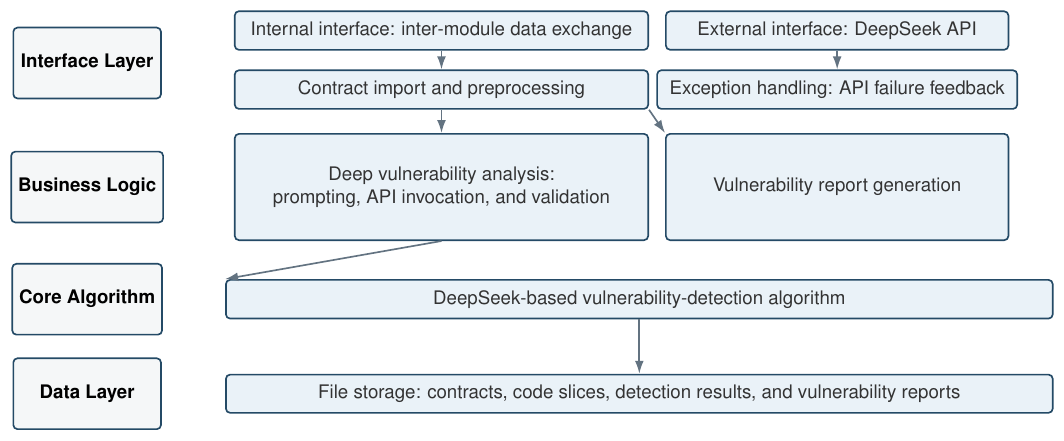}
  \caption{Layered architecture.}
\end{subfigure}
\caption{Overall framework of the proposed vulnerability-detection method.}
\label{fig:overall-framework}
\end{figure}

We organize the system into contract preprocessing, DeepSeek-based analysis, and result generation. Contracts and vulnerability-category labels are loaded from a CSV file. The preprocessing stage locates candidate statements, extracts line-numbered code slices, and writes normalized inputs. The analysis stage constructs a knowledge-guided prompt and invokes the official DeepSeek API, while the result stage parses the fixed response fields and aggregates contract-level labels for statistical analysis. Figure~\ref{fig:overall-framework} presents the functional modules and layered software architecture.

\subsection{Vulnerability-Focused Code Slicing}

\begin{figure}[!t]
\centering
\begin{subfigure}[t]{0.94\textwidth}
  \centering
  \includegraphics[width=0.78\linewidth]{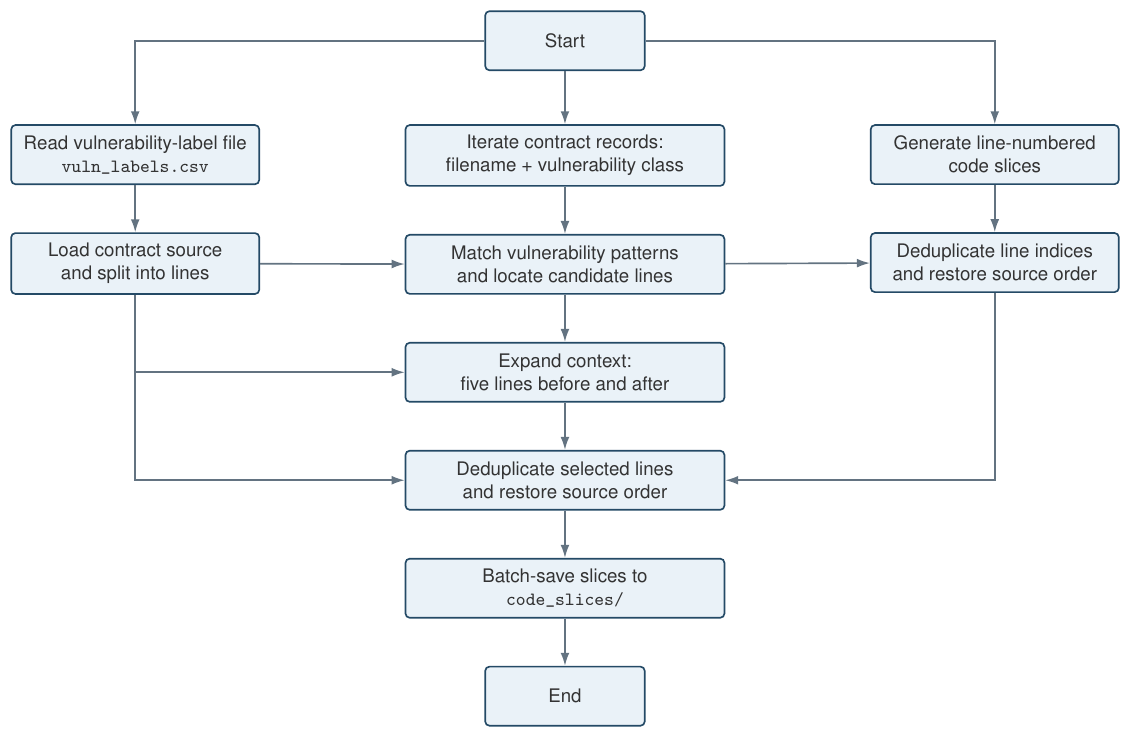}
  \caption{Preprocessing flow.}
\end{subfigure}

\vspace{0.7em}
\begin{subfigure}[t]{0.94\textwidth}
  \centering
  \includegraphics[width=0.82\linewidth]{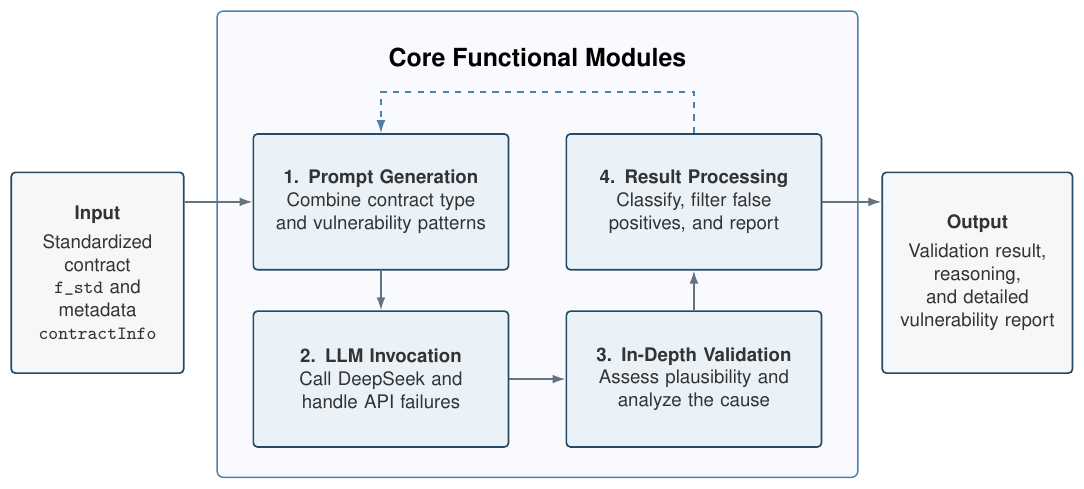}
  \caption{DeepSeek analysis flow.}
\end{subfigure}
\caption{Contract preprocessing and knowledge-guided DeepSeek analysis.}
\label{fig:slicing-analysis}
\end{figure}

\begin{figure}[!t]
\centering
\includegraphics[height=0.50\textheight]{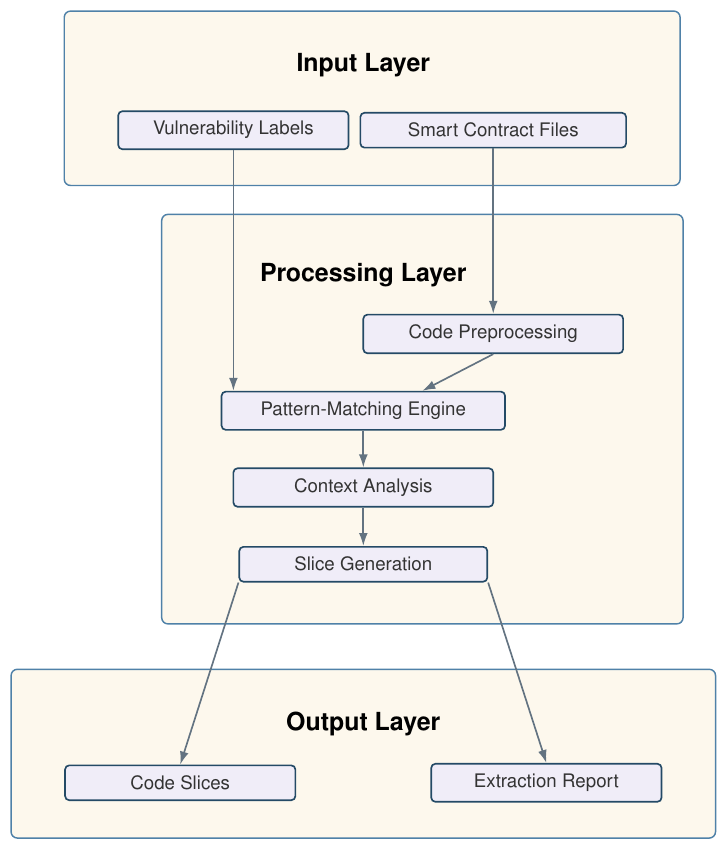}
\caption{Architecture of the code-slice extraction module.}
\label{fig:slice-architecture}
\end{figure}

The slicing stage uses class-specific regular expressions as a high-recall filter. When a pattern matches, the algorithm first creates a local window and then searches backward for a likely function start and forward for a balanced closing brace. Figures~\ref{fig:slicing-analysis} and~\ref{fig:slice-architecture} present the preprocessing, DeepSeek-analysis, and slice-architecture workflows as English vector diagrams. Algorithm~\ref{alg:slicing} formalizes the code-slicing procedure without reproducing the complete Python implementation. The initial window is five lines on each side, while the forward scan may examine up to fifty lines to find the end of a function. Table~\ref{tab:patterns} lists representative patterns used by the preprocessing module. These expressions are candidate locators rather than vulnerability proofs.

\begin{algorithm}[H]
\caption{Structure-aware extraction of vulnerability-focused code slices}
\label{alg:slicing}
\begin{algorithmic}[1]
\Require Contract lines $C$, vulnerability class $v$, pattern library $P_v$
\Ensure Set of selected line indices $S$
\State $S \gets \emptyset$
\For{$i \gets 1$ to $|C|$}
  \If{$C_i$ is ignorable}
    \State \textbf{continue}
  \EndIf
  \If{any pattern in $P_v$ matches $C_i$}
    \State $l \gets \max(1,i-5)$; $r \gets \min(|C|,i+5)$
    \State Search backward from $i$ for a function declaration or opening block; update $l$
    \State Search forward from $i$ while balancing braces; update $r$
    \State $S \gets S \cup \{l,\ldots,r\}$
  \EndIf
\EndFor
\State \Return $S$
\end{algorithmic}
\end{algorithm}

\begin{table}[H]
\centering
\caption{Representative patterns used for candidate-line extraction.}
\label{tab:patterns}
\small
\begin{tabularx}{\textwidth}{>{\raggedright\arraybackslash}p{0.22\textwidth} X}
\toprule
Class & Representative patterns \\
\midrule
Reentrancy & \texttt{transfer(}, \texttt{transferFrom(}, \texttt{emergencyWithdraw}, \texttt{withdrawToAddress}, purchase and withdrawal function names. \\
Integer overflow/underflow & Compound assignments such as \texttt{+=}, \texttt{-=}, and \texttt{*=}; direct binary arithmetic; exponentiation expressions. \\
Timestamp dependence & \texttt{block.timestamp}, \texttt{block.number}, \texttt{now}; timestamp comparisons in \texttt{require}, \texttt{if}, and \texttt{assert}. \\
\bottomrule
\end{tabularx}
\end{table}

\subsection{Knowledge-Guided DeepSeek Analysis}

For each slice, the prompt supplies an ERC-721-oriented knowledge base, the line-numbered code, a closed label set, and explicit rejection rules. DeepSeek must choose at most one of the three target classes, avoid speculation about unavailable code, return \texttt{None} when the evidence is insufficient, and identify a suspicious line range. Prompt~\ref{lst:prompt} summarizes the prompt structure used to constrain DeepSeek. Because the complete external knowledge base is not included, the template records the input structure, decision constraints, and output schema rather than the full experimental prompt.

\noindent\begin{minipage}{\textwidth}
\begin{lstlisting}[caption={Abridged prompt template for knowledge-guided DeepSeek analysis.},label={lst:prompt}]
System role:
  Ethereum NFT smart-contract vulnerability analyst

Allowed output label:
  Reentrancy | Integer overflow/underflow |
  Timestamp dependence | None

Domain knowledge:
  <ERC-721_DOMAIN_KNOWLEDGE>

Input:
  <LINE-NUMBERED_CODE_SLICE>

Decision constraints:
1. Analyze only the logic contained in the supplied slice.
2. Do not assume behavior that is not shown in the input.
3. Return None when the available evidence is insufficient.
4. Report suspicious locations using the format Lines X-Y.

Required output schema:
Vulnerability Type: <LABEL>
Suspicious Code Location: <LINES X-Y>
Vulnerability Cause: <RATIONALE>
\end{lstlisting}
\end{minipage}

The official DeepSeek API receives the prompt and returns a text response. The implementation handles request failures and timeouts before parsing the three required fields. Batch processing iterates over filenames containing \texttt{slice.sol}, appends results to a detection file, and converts the parsed records into a data frame. We describe these routine engineering operations in prose rather than reproduce the complete implementation.

\section{Evaluation}

\subsection{Experimental Setup}

The experiments ran on Windows 11 with an Intel Core i7-9750H processor at 2.60~GHz and 32~GB of memory. The implementation used Python 3.11 for API communication, structured aggregation, plotting, regular-expression processing, and file handling. The exact DeepSeek model identifier, decoding temperature, token limit, API version, retry policy, and complete knowledge-base text were not recorded, which limits reproducibility. The evaluation comprised three parts: checks on selected samples, automated processing of 450 NFT contract samples, and two ablations. The selected samples came from real NFT projects or public vulnerability datasets; however, contract addresses, dataset identifiers, independent expert labels, and an annotation protocol were not recorded. We therefore describe the primary metric as a positive-label rate rather than independently verified accuracy.

\subsection{Selected-Sample Checks}

Table~\ref{tab:single-samples} summarizes the structured outputs for three selected vulnerability samples. Each row reports the evidence and interpretation for one sample.

\begin{table}[H]
\centering
\caption{Reported outputs for the three selected vulnerability samples.}
\label{tab:single-samples}
\small
\begin{tabularx}{\textwidth}{>{\raggedright\arraybackslash}p{0.18\textwidth} X X}
\toprule
Sample & Reported evidence & Reported interpretation \\
\midrule
Reentrancy & An external transfer occurs before the expected state update and without an effective reentrancy lock. & Reentrancy vulnerability. \\
Integer overflow/underflow & The contract uses custom arithmetic associated with a Solidity version earlier than 0.8.0. & Potential unchecked integer arithmetic. \\
Timestamp dependence & Fundraising checks rely on \texttt{now} or equivalent timestamp conditions in multiple functions. & Timestamp-dependent control logic. \\
\bottomrule
\end{tabularx}
\end{table}

These examples illustrate that the structured response contains a class, location, and rationale. Because the samples were selected and no independent adjudication was conducted, they are functional demonstrations rather than a general accuracy estimate.

\subsection{Batch Results}

The batch evaluation processed 450 NFT contract slices and produced 437 positive labels and 13 \texttt{None} labels. This corresponds to $437/450=97.11\%$, rounded to 97.1\%. The class distribution contains 155 reentrancy labels, 141 integer overflow or underflow labels, 141 timestamp-dependence labels, and 13 \texttt{None} labels. Slice extraction required 13.74s in total, approximately 0.03s per contract. An average detection request required about 3.8s, while the complete batch required 34min, corresponding to 4.53s per sample end to end. The difference suggests that the end-to-end batch time includes API and orchestration overhead.

\subsection{Ablation Study}

\begin{figure}[H]
\centering
\begin{subfigure}[t]{0.48\textwidth}
  \centering
  \includegraphics[width=\linewidth]{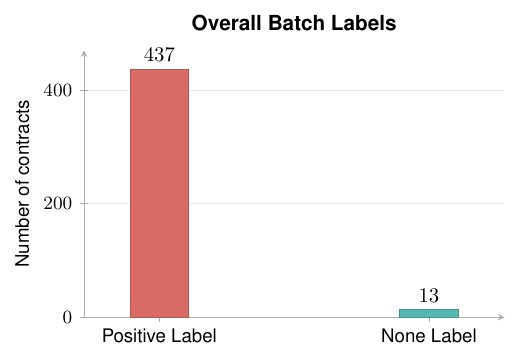}
  \caption{Overall batch labels.}
\end{subfigure}
\hfill
\begin{subfigure}[t]{0.48\textwidth}
  \centering
  \includegraphics[width=\linewidth]{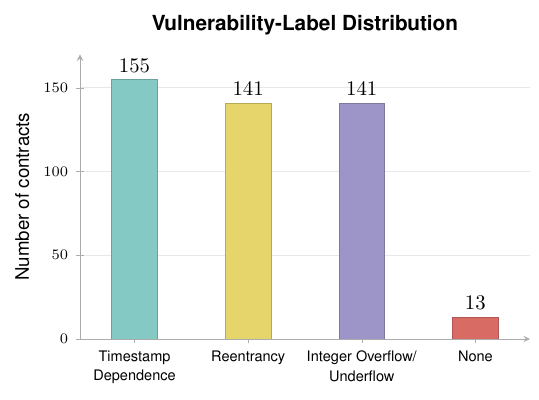}
  \caption{Vulnerability-label distribution.}
\end{subfigure}

\vspace{0.8em}
\begin{subfigure}[t]{0.48\textwidth}
  \centering
  \includegraphics[width=\linewidth]{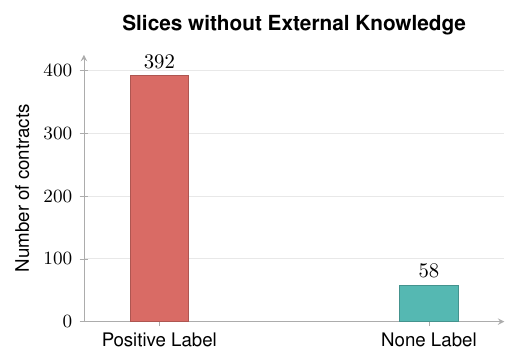}
  \caption{Slices without external knowledge.}
\end{subfigure}
\hfill
\begin{subfigure}[t]{0.48\textwidth}
  \centering
  \includegraphics[width=\linewidth]{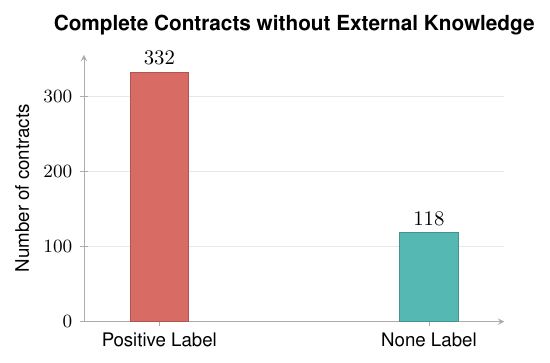}
  \caption{Complete contracts without external knowledge.}
\end{subfigure}
\caption{Detection results and ablation comparison.}
\label{fig:results}
\end{figure}

Two ablations examine the roles of domain knowledge and focused context. When core code slices were analyzed without the external knowledge base, 392 of 450 samples received positive labels and 58 received \texttt{None} labels, giving 87.11\%. When complete contracts were analyzed without the knowledge base, 332 samples received positive labels and 118 received \texttt{None} labels, giving 73.78\%. The complete-contract run required 37min, or 4.93s per contract. Table~\ref{tab:ablation-results} reports the numerical results, and Figure~\ref{fig:results} visualizes the batch and ablation outcomes.

\begin{table}[H]
\centering
\caption{Reported results for the full and ablation configurations.}
\label{tab:ablation-results}
\small
\begin{tabularx}{\textwidth}{X r r r}
\toprule
Configuration & Positive & \texttt{None} & Reported rate \\
\midrule
Code slices with external knowledge & 437 & 13 & 97.11\% \\
Code slices without external knowledge & 392 & 58 & 87.11\% \\
Complete contracts without external knowledge & 332 & 118 & 73.78\% \\
\bottomrule
\end{tabularx}
\end{table}

Adding external knowledge to sliced inputs increased the reported positive-label rate by 10.00 percentage points, from 87.11\% to 97.11\%. Under the no-knowledge condition, replacing complete contracts with code slices increased the rate by 13.33 percentage points, from 73.78\% to 87.11\%. These controlled comparisons indicate that focused context and knowledge constraints affect the reported outputs. However, the metric cannot establish precision or recall without ground-truth labels.

\section{Discussion and Limitations}

The method combines two forms of task restriction. Code slicing removes much of the contract content unrelated to a candidate vulnerability, while the knowledge base and prompt define the permitted labels and decision rules. These constraints make the LLM response easier to parse and limit unrestricted explanations. The method modifies preprocessing and prompting without training a new model.

Several limitations constrain the conclusions. First, the evaluation does not include independently annotated ground truth, a confusion matrix, per-class precision or recall, or inter-annotator agreement. A positive-label rate must not be interpreted as classification accuracy. Second, the vulnerability-category labels used during preprocessing were generated semi-automatically, which may reveal expected classes to the slicing procedure. Third, the exact DeepSeek configuration and full knowledge base were not recorded, limiting reproducibility and analysis of output stability.

The detector covers only three vulnerability classes. Regular expressions may miss semantically equivalent implementations or match harmless identifiers, and local slices can omit inherited modifiers, storage invariants, cross-function calls, or cross-contract behavior. Timestamp-related patterns may also produce false positives because legitimate contracts can use timestamps for noncritical purposes. Future work should publish the dataset and scripts, freeze the prompt and model configuration, establish expert-reviewed labels, compare against static and LLM baselines, and report per-class precision, recall, F1 score, and confidence intervals.

\section{Conclusion}

We present a knowledge-guided method for detecting reentrancy, integer overflow or underflow, and timestamp dependence in Ethereum NFT smart contracts. The pipeline uses regular-expression screening and a structure-aware context window to extract line-numbered code slices, then constrains DeepSeek with ERC-721 domain knowledge and a fixed response schema. On 450 samples, the full configuration produced 437 positive labels, corresponding to a 97.1\% positive-label rate. The reported rate decreased to 87.11\% without the external knowledge base and to 73.78\% when complete contracts were analyzed without that knowledge. These results indicate that code slicing and domain constraints materially change the detector output.

\section*{Acknowledgments}

AI-based tools are used for language polishing during manuscript preparation.

\bibliographystyle{unsrt}
\bibliography{references_v2}

\end{document}